# Hybrid Topological Defects in Ferroelectric Nematic Fluids

Shengzhu Yi[1,2], Chao Zhou[1], Zening Hong[1], Xinming Chu[1], Zhongjie Ma[1], Mingjun Huang[3,4], Satoshi Aya[3,4], Rui Zhang[2,5], Qi-Huo Wei[1,6]*

[1]State Key Laboratory of Quantum Functional Materials, Department of Mechanical and Energy Engineering, Southern University of Science and Technology, Shenzhen, 518055, China
[2]Department of Physics, The Hong Kong University of Science and Technology, Clear Water Bay, Hong Kong, China
[3]South China Advanced Institute for Soft Matter Science and Technology (AISMST), School of Emergent Soft Matter, South China University of Technology, Guangzhou 510640, China
[4]Guangdong Provincial Key Laboratory of Functional and Intelligent Hybrid Materials and Devices, South China University of Technology, Guangzhou 510640, China
[5]State Key Laboratory of Displays and Opto-electronics, The Hong Kong University of Science and Technology, 999077, Hong Kong, China
[6]Center for Complex Flows and Soft Matter Research, Southern University of Science and Technology, Shenzhen 518055, China

We investigate the evolution of topological defects in polar fluids driven by discrete inversion symmetry-breaking across the nematic-to-ferroelectric nematic phase transition. Using photopatterned surface alignment to prescribe well-defined initial defect configurations in the nematic phase, we track their metamorphosis upon entering the ferroelectric phase. Through systematic comparisons between experiments and numerical simulations, we demonstrate that the resulting polar structures are generically hybrid, intertwining topological configurations of different dimensionalities. Specifically, we identify three basic hybrid defects evolving from precursor disclinations of net charge 0 and +1 (via pairs), and +1 (via single defects): domain walls terminated by surface disclinations, monopole-decorated domain walls, and meron-mediated boojum and monopole complexes. We further demonstrate that these stable hybrid polar defects are ubiquitous in the ferroelectric nematic phase, providing the basic motifs needed to understand, predict, and engineer complex topological textures in polar fluids.

Topological defects, singular configurations in ordered media, are ubiquitous hallmarks of broken symmetry [1]. These defects emerge across a vast range of scales, from the nascent universe [2,3] to diverse condensed matter systems like superfluid $^3$He [4–6], ferroic materials [7,8], and liquid crystals [9–11]. The emergence and proliferation dynamics of topological defects during spontaneous symmetry-breaking transitions have been described in field theories by Kibble and Zurek for cosmic defects like strings and domain walls [12]. This theoretical framework was subsequently extended to predict that sequential phase transitions, involving the breaking of both continuous and discrete symmetries, could yield hybrid structures where defects of different dimensionalities intertwine [12,13].

The universality of symmetry-breaking concepts, coupled with the difficulty of directly detecting cosmological structures, has driven the search for analogs in condensed matter systems such as liquid crystals [14,15]. Traditional liquid crystals, composed of rod-like molecules, can align along a common direction called the director (**n**), forming the nematic phase. Upon cooling, they transition from an isotropic phase (I) with full symmetry to the nematic phase (N) that retains rotational symmetry [SO(2)] only around the director **n** and the head-to-tail symmetry ($\mathbb{Z}_2$) of **n**, thereby breaking continuous rotational symmetry [10]. These conventional liquid crystals have been an exceptional laboratory system for both validating the Kibble-Zurek mechanism [14,15] and exploring intriguing properties of topological defects [16–18]. So far, hybrid topological defects have been observed only in superfluid $^3$He [19] and as two-dimensional variants in suspended tilted smectic films [20–22].

A newly discovered class of polar liquid crystals exhibits an additional transition from the apolar nematic phase to a polar nematic ($N_F$) phase characterized by a spontaneous polarization, **P** [23–26]. This secondary transition breaks discrete inversion symmetry ($\mathbb{Z}_2$) and reduces the symmetry to SO(2), providing a physical realization of the symmetry-breaking cascade envisioned by Kibble. Formations of hybrid topological defects in the $N_F$ phase have been proposed [26,27], but direct evidence remains absent. The N–$N_F$ transition is known to host a rich variety of topological defects and associated dynamical phenomena [28–37]; however, the hybrid nature of these polar defects and their connections to inversion symmetry-breaking remain to be fully elucidated. Here we report on the first direct experimental and numerical observation of these hybrid polar defects, identifying three configurations and establishing the topological principles behind the defect evolution across the N–$N_F$ transition.

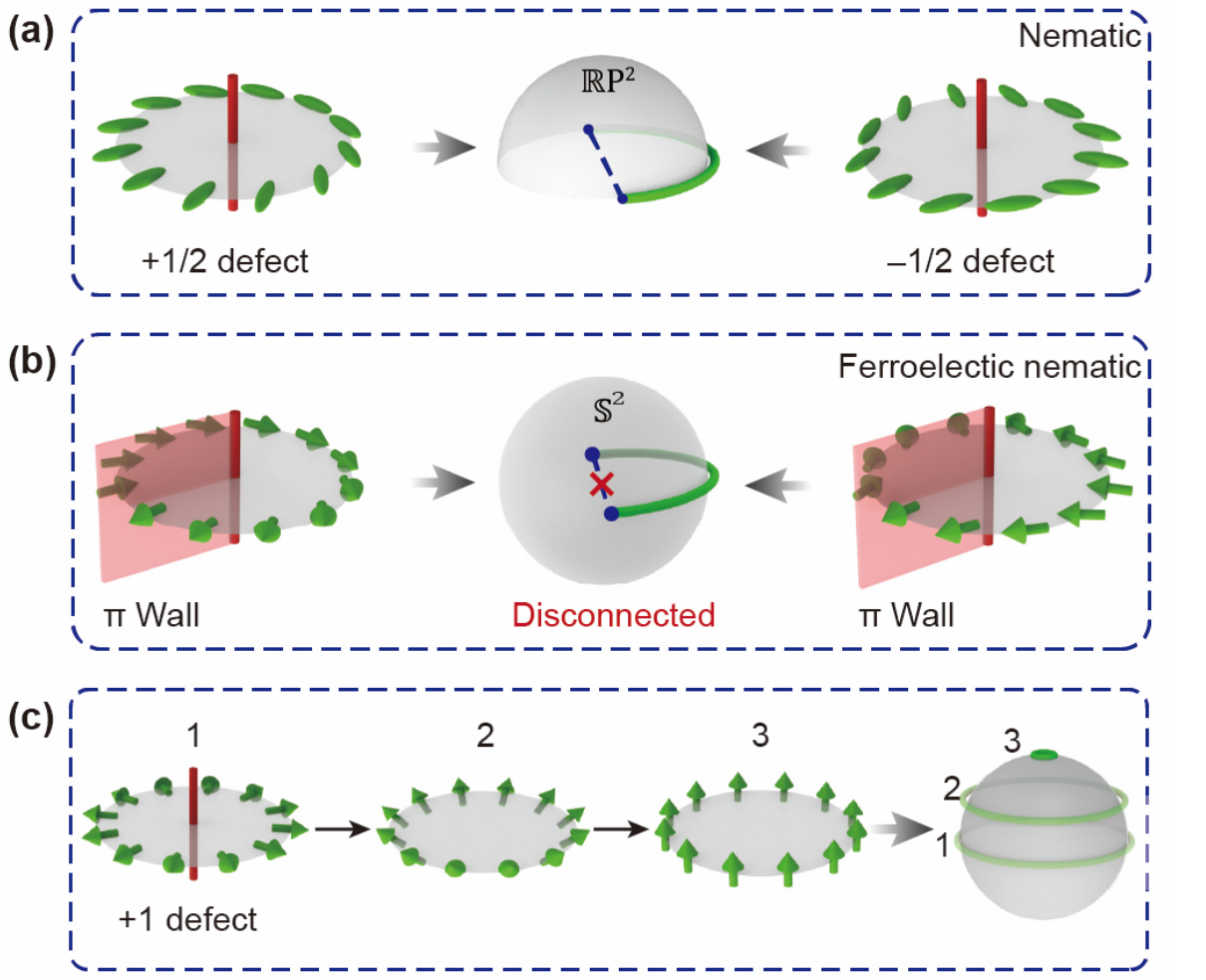

FIG. 1. (a) Director fields around ±1/2 disclinations and their mapping onto the ground-state manifold in the N phase. (b) Polarization fields around ±1/2 defects and their mapping onto the ground-state manifold in the $N_F$ phase. (c) Topological instability of integer disclinations.

*Symmetry, Topology and Experiments*—Inversion symmetry-breaking changes the order-parameter space, or the ground-state manifold. In the N phase, the ground-state manifold is the real projective plane ($\mathbb{R}P^2$), corresponding to a sphere with antipodal points identified by the head-tail symmetry of the director **n**. In the $N_F$ phase, this degeneracy is lifted by the emergence of polar order, transforming the manifold into the sphere ($\mathbb{S}^2$). This change fundamentally alters the homotopy groups that determine the topological stability of defects.

In the N phase, the fundamental group $\pi_1(\mathbb{R}P^2) = \mathbb{Z}_2$ guarantees that half-integer disclinations are topologically stable: the director field on a closed loop encircling the defect maps to a continuous path connecting antipodal points on $\mathbb{R}P^2$, which cannot be continuously contracted to a point [Fig. 1(a)]. Integer disclinations, by contrast, are not topologically protected; their mappings form closed loops on $\mathbb{R}P^2$ that can continuously contract to a point, enabling the singular core to escape into the third dimension. They therefore typically dissociate into a stable pair of half-integer disclinations in the N phase [38].

In the $N_F$ phase, $\pi_1(\mathbb{S}^2) = 0$, and isolated half-integer disclinations are forbidden; their corresponding mapping sweeps only half of a great circle on $\mathbb{S}^2$ and cannot close. They can exist only if terminated by a domain wall or bound to another topological structure, forcing the coexistence of defects of different dimensionalities [Fig. 1(b)]. Integer disclinations remain unprotected in the polar phase for the same reason as in the N phase [Fig. 1(c)], but now they can no longer dissociate into half-integer disclinations, which are themselves forbidden as isolated objects.

To investigate how this transformation of the ground-state manifold drives defect evolution, we consider liquid crystal cells in which identical planar director patterns are imprinted on both confining surfaces. The director field is apolar and confined to the surface plane, parameterized as: $\mathbf{n}(x, y) = [\cos\theta(x, y), \sin\theta(x, y), 0]$, where the orientation angle $\theta(x, y)$ is given by:

$$\theta(x, y) = \sum_i q_i \tan^{-1}[(y - y_{0i})/(x - x_{0i})] + \alpha \qquad (1)$$

with $q_i$ denoting the topological charge of the $i$-th defect located at $(x_{0i}, y_{0i})$, and $\alpha$ a constant phase shift. The net topological charge within the $xy$-plane is therefore: $Q = \sum_i q_i$.

We focus on three basic configurations of disclinations: a (+1/2, −1/2) disclination pair with net charge $Q = 0$, a pair of +1/2 disclinations with $Q = +1$, and a single isolated +1 disclination. These configurations serve as minimal building blocks to elucidate how the defect evolution across the N–$N_F$ transition is jointly governed by two distinct mechanisms: topological constraint, which prohibits isolated half-integer defects in the polar phase, and topological instability, which destabilizes isolated integer disclinations.

The experimental cell consists of two flat glass substrates with inner surfaces coated with the photoalignment material azo-dye SD1. The director fields prescribed by Eq. (1) are imprinted into the SD1 layers using a home-built photopatterning system [Fig. S1] [39–41]. The cells are subsequently filled with RM734, which exhibits the phase sequence I-187°C-N-133°C-$N_F$-84°C-Crystal. To avoid the ground state with spontaneous polarization twist across the cell, known to prevail in thicker cells to minimize electrostatic energy [51,52], we restrict the cell thickness to < 3 μm.

*Transformation of a ±1/2 Disclination Pair into a Domain Wall Terminated by Surface Disclinations*—The first case has $q_{1,2} = \pm 1/2$, $\alpha = \pi/2$ [Eq. (1)], yielding a net topological charge $Q = 0$. In this configuration, the near-field director aligns parallel to the line connecting the two defects but gradually rotates toward the periphery to become perpendicular to it [Fig. 2(a)]. This configuration was proposed by Lavrentovich as an intriguing pathway to forming a π domain wall bounded by vertical half-integer disclinations, where the polarizations beside the domain wall point in opposite directions [26]. In the 2D XY model, this director field manifests as a string defect connecting two half-integer vortices, which provides a pathway for topologically forbidden polar defects to be excited in pairs [53].

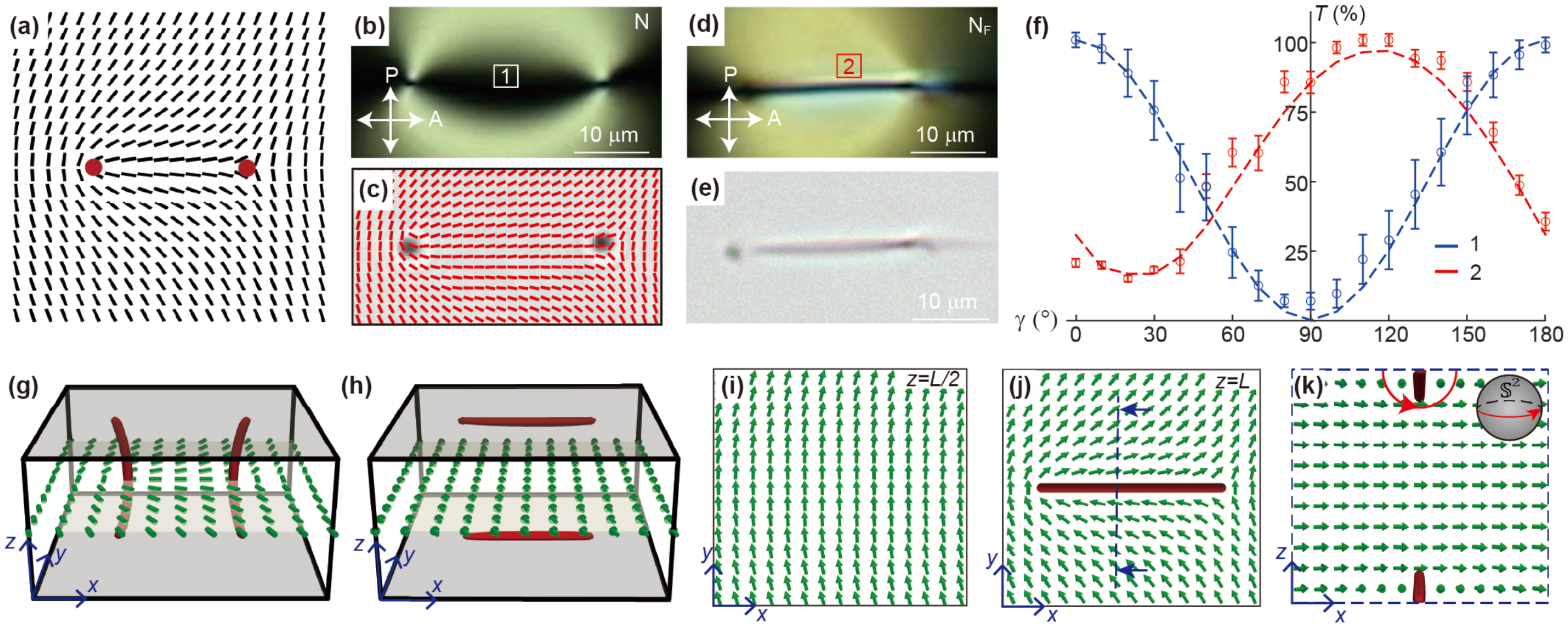

FIG. 2. Transformation of a ±1/2 disclination pair into a domain wall. (a) Director field of the ±1/2 defect pair. (b, c) Polarizing optical image (b) and measured director field (c) in the N phase; the background in (c) is a bright-field image. (d, e) Polarizing optical (d) and bright-field (e) images in the $N_F$ phase. The cell thickness is $L$=1.5 μm, and the material used is RM734. (f) Optical transmittance measured in regions 1 (b) and 2 (d) as a function of the analyzer-polarizer angle $\gamma$; dashed curves show calculated transmittance from the simulated orientation field. (g, h) Calculated defect configurations in the N (g) and $N_F$ (h) phases. (i, j) Calculated polarizations within $xy$-planes at $z$ = 0.5$L$ (i) and $z$ = $L$ (j). (k) Calculated polarizations in the cross-sectional plane indicated by the dashed line in (j). The two blue arrows in (j) indicate the viewing direction. Dark brown regions in (g-k) denote areas where the order parameters $s$ and $p$ fall below 0.45.

The experimental results for the N phase are shown in Figs. 2(b) and 2(c), where crossed polarized optical microscopy and the director field measured via the PolScope technique [54] reveal that the printed liquid crystal's director field precisely matches the design, containing two half-integer disclinations. We then performed numerical simulations based on the Landau–de Gennes theory [41,55], which revealed two possible disclination configurations, depending on cell thickness: (i) in thin cells, two vertical disclination lines bridging the point defects that bend towards each other [Fig. 2(g)], and (ii) in thick cells, two horizontal disclination lines connecting the +1/2 and −1/2 defects (Fig. S2 [41]). Our experimental condition corresponds to the first case.

Upon entering the $N_F$ phase, we observe two dark lines extending from the +1/2 and −1/2 disclinations and merging into a continuous domain wall, and the black brushes between two defects vanishing (Movie S1 [41]). We measured the optical transmittance ($T$) as a function of the angle ($\gamma$) between the polarizer and analyzer, with the polarizer oriented parallel to the local director imprinted on the confining surfaces. In contrast to a uniform cell where $T(\gamma) \propto \cos^2\gamma$, the optical transmittance in the region denoted by the square in Fig. 2(d) is not extinguished. When the polarizers are uncrossed, an asymmetric change in transmittance and domain color can be observed (Figs. 2(f) and S3 [41]). These optical features suggest that the director is twisted along the cell normal.

We simulate the polarization field based on an extended Oseen–Frank free-energy functional [56] and reveal that the defect structure in the $N_F$ phase consists of line defects bound to the confining surfaces [Fig. 2(h)]. Near these defects, the polarization undergoes a sharp $\pi$ reversal, whereas at the cell midplane it remains uniformly aligned [Fig. 2(i, j)], resulting in a polarization twist along the cell normal. Optical transmittance, calculated from the simulated polarization field using the Jones matrix method, shows quantitative agreement with the experimental results [Fig. 2(f)].

The line defects confined to the surfaces connect a +1/2 and a −1/2 point defect [Fig. 2(j)], forming string defects as originally proposed in the 2D XY model [53]. Mapping the polarization along a semicircular path around the line defect traces a half-circle connecting antipodal points on the $\mathbb{S}^2$ manifold, confirming their identity as half-integer surface disclinations [Fig. 2(k)]. These surface disclinations remain anchored to the confining substrates, in contrast to the originally proposed $\pi$ domain wall terminated by a pair of vertical half-integer disclinations [26]. We attribute the absence of vertical disclinations to energetic considerations:

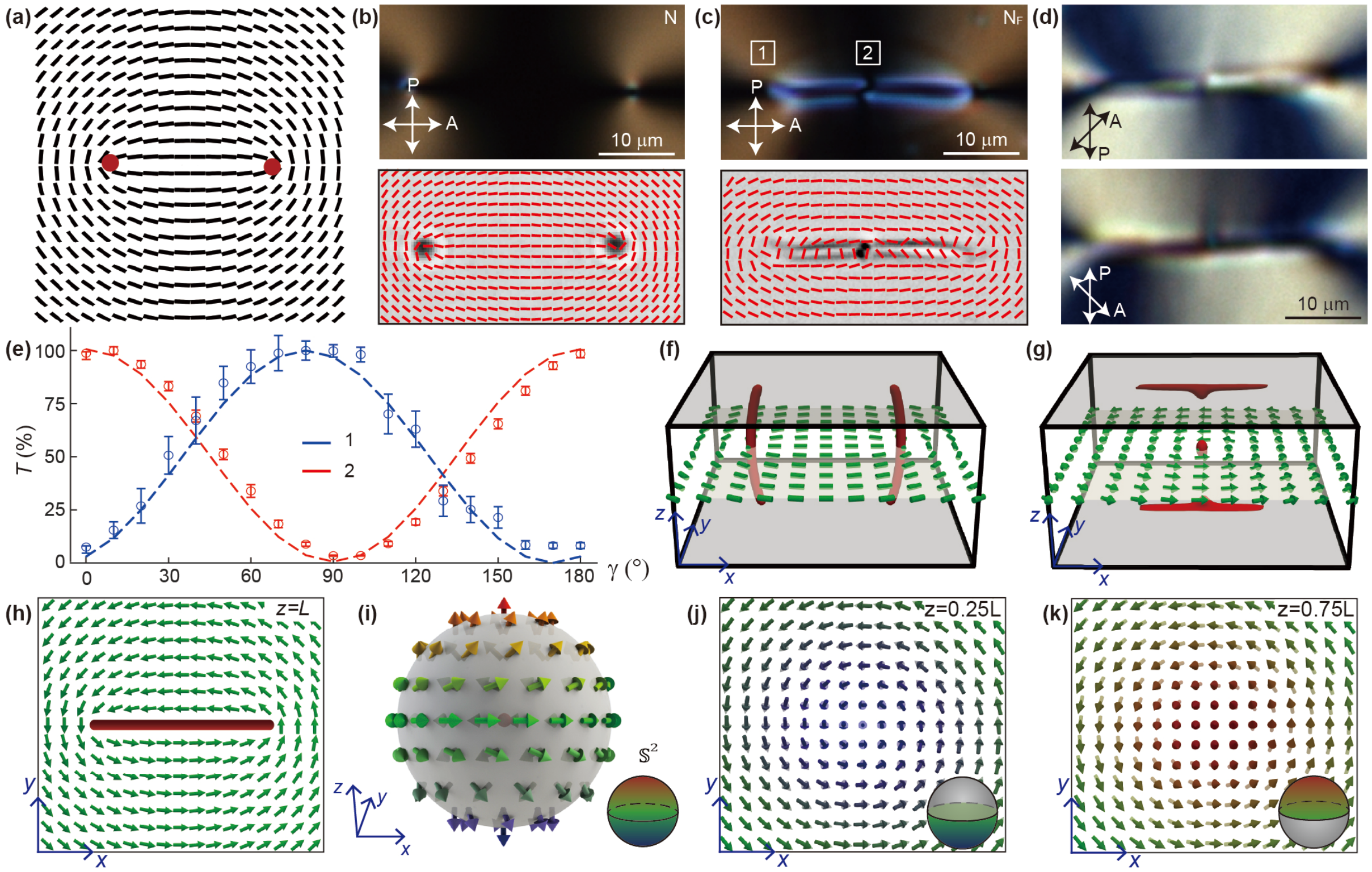

FIG. 3. Transformation of a +1/2 disclination pair into a monopole-decorated domain wall. (a) Director field of the +1/2 defect pair. (b, c) Polarizing optical images (top) and measured director fields (bottom) in the N phase (b) and $N_F$ phase (c). (d) Polarizing optical images of the defect with the analyzer oriented at 45° and 135° relative to the polarizer. Cell thickness: $L$ = 2.0 μm. The material used is RM734. (e) Optical transmittance in regions 1 and 2 of (c) as a function of analyzer-polarizer angle γ; dashed curves show calculated transmittance from the simulated director field. (f, g) Simulated defect configurations in the N (f) and $N_F$ (g) phases. (h-k) Calculated polarizations near the top substrate (h), on a spherical surface encompassing the monopole and its mapping into the ground-state manifold (i), and within $xy$-planes at $z = 0.25L$ (j) and $z = 0.75L$ (k). Dark brown regions in (f, g, h) denote areas where the order parameters $s$ and $p$ fall below 0.45. The color of the arrows and of the ground state manifold $\mathbb{S}^2$ in (i-k) represents the angle between **P** and the $z$ axis.

forming two surface disclinations is less costly than creating a singular domain wall accompanied by two vertical disclinations.

*Transformation of a pair of* +1/2 *Disclinations into a Monopole-Decorated Domain Wall*—The transformation of the ±1/2 pair into two surface disclinations relies on its vanishing net charge, which allows the midplane to become defect-free. We next extend this construction to a basic counterpart with nonzero topological charge, a pair of +1/2 defects. The associated director field, described by Eq. (1), is characterized by $q_{1,2} = +1/2$ and $\alpha = \pi/2$, yielding $Q = +1$. The defects are arranged in a head-to-head configuration, such that the director along the line connecting them is parallel to the line [Fig. 3(a)].

In the N phase, polarized optical microscopy reveals two dark brushes emanating from the defect centers, and PolScope measurements confirm a director field that closely matches the imprinted pattern [Fig. 3(b)]. Numerical simulations show that these defects correspond to two vertical disclinations linking the point defects on the opposing confining surfaces [Fig. 3(f)]. Upon entering the $N_F$ phase, the dark brushes are retained, as shown in Fig. 3(c). The key difference is the appearance of a point defect at the center of a black domain wall, most evident under bright-field imaging (Fig. 3(c) and Movie S2 [41]). This contrasts with the smooth domain wall observed in the ±1/2 defect pair. Optical transmittances measured with the polarizer approximately aligned with the imprinted local director exhibit the expected $T(\gamma) \propto \cos^2\gamma$ dependence for regions 1 and 2 [Fig. 3(d) and 3(e)], indicating the absence of director twist across the cell. PolScope imaging further confirms that the director near the domain wall remains largely parallel to it [Fig. 3(c)]. We note that measurement accuracy of director orientation within the wall region is limited by the complex 3D nature of the local director field.

Numerical calculations show that the defect structure in the $N_F$ phase consists of two surface disclinations and a point defect at the cell midplane [Fig. 3(g)]. To characterize the point defect, we map the polarization on a spherical surface surrounding its core onto the ground-state manifold. This mapping completely covers the $\mathbb{S}^2$ surface, indicating that the defect is a monopole [Fig. 3(i)]. To determine the topological charge $Q_{\mathrm{mp}}$ carried by the monopole, we integrate the polarization unit vector $\mathbf{p}$ over an enclosing sphere $\sigma$ with spherical polar coordinates $u_1$ and $u_2$: $Q_{\mathrm{mp}} = \int_\sigma (\mathbf{p} \cdot \partial_{u_1}\mathbf{p} \times \partial_{u_2}\mathbf{p}) du_1 du_2 / 4\pi$. The result confirms that the topological charge of the monopole is +1.

In any $xy$-plane between the monopole and the surface disclinations, the polarization lies in the plane at the periphery but becomes vertical near the center [Fig. 3(j, k)]. Mapping these configurations onto the $\mathbb{S}^2$ manifold shows that regions above and below the midplane cover the upper and lower hemispheres, respectively, indicating that the in-plane textures correspond to merons (half-Skyrmions).

*Transformation of a +1 Disclination into Meron-Mediated Boojum and Monopole Complexes*—The surface disclinations associated with a ±1/2 pair carry no net charge and can continuously annihilate into a defect-free texture. In contrast, those associated with a +1/2 defect pair carry a net charge $Q = +1$, and can deform into a +1 point defect as two +1/2 defects approach and merge. This coalescence limit naturally motivates us to examine the integer defect, described by Eq. (1) with $q_1 = +1$ and $\alpha = 0$. In this configuration, the radial director exhibits pure splay deformations [Fig. 4(a)].

In the N phase, the optical texture reveals two defect cores, each associated with two black brushes, characteristic of half-integer strength. Our simulations verify the splitting into two +1/2 disclinations, highlighting both the director fields and the reduced order parameter at the cores of the two +1/2 defects in the cell's midplane [Fig. 4(d)].

As the liquid crystal enters the $N_F$ phase, the optical textures of these two +1/2 defects collapse into a single defect characterized by four dark brushes (Fig. 4(c) and Movie S3 [41]). These dark brushes form a distinctive pinwheel structure, which adopts either a left- or right-handed twist with equal probability (Fig. S4 [41]). By analyzing the response of this spiral texture to an in-plane electric field, we determined that the polarization points outward (Fig. S5 [41]), indicating that $\mathbf{P}$ is aligned parallel to $\mathbf{n}(\nabla \cdot \mathbf{n})$. This specific polarization can be attributed to the flexoelectric effect. RM734 molecules possess a pear shape, with polarization directed toward the larger molecular end. This shape promotes close packing that favors splay deformations, giving rise to a positive flexoelectric coupling coefficient ( $\lambda$ ). Accordingly, the flexoelectric contribution to the free energy, $F_{\text{Flexo}} = -\lambda \int dr^3 \mathbf{n}(\nabla \cdot \mathbf{n}) \cdot \mathbf{P}$, is minimized when the polarization $\mathbf{P}$ aligns parallel to the splay vector $\mathbf{n}(\nabla \cdot \mathbf{n})$ [52,57,58].

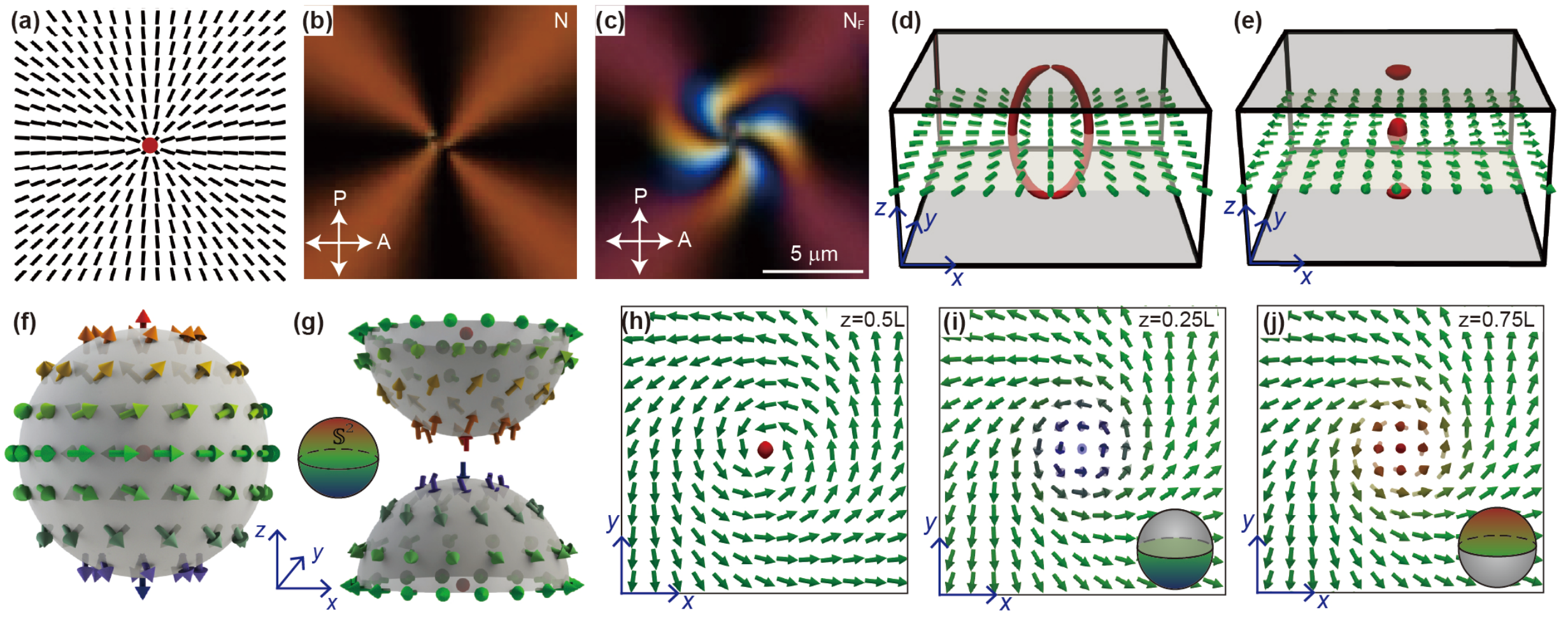

FIG. 4. Transformation of a radial +1 disclination into meron-mediated boojum and monopole complexes. (a) Director field of the +1 defect. (b, c) Polarizing optical images of a +1 defect in the N (b) and the $N_F$ (c) phases. Cell thickness: $L$ = 2.6 μm. The material used is RM734. (d, e) Simulated defect structures in the N (d) and the $N_F$ (e) phases. (f) Calculated polarization field on a spherical surface encompassing the monopole. (g) Calculated polarization fields on hemispherical surfaces encompassing the boojums at the top and bottom substrates. (h-j) Calculated polarization fields in $xy$-planes at $z$ = 0.5$L$ (h), $z$ = 0.25$L$ (i), and $z$ = 0.75$L$ (j). Dark brown regions in (d, e) denote areas where the order parameters $s$ and $p$ fall below 0.45. The color of the arrows and of the ground state manifold $\mathbb{S}^2$ in (f-j) represents the polar angle between $\mathbf{P}$ and the z axis.

Numerically calculated polarization fields reveal that the polar defect consists of a point defect in the cell's midplane accompanied by two additional point defects at the confining surfaces [Fig. 4(e)]. To determine the topological nature of these point defects, we map the polarization field onto the ground-state manifold. Projecting the polarization on a spherical surface enclosing the central defect yields complete coverage of the $\mathbb{S}^2$ manifold, identifying it as a monopole with charge $Q_{\text{mp}} = +1$ [Fig. 4(f)]. In contrast, mapping the polarization on hemispherical surfaces centered at the surface defect cores produces coverage restricted to a single hemisphere of the manifold [Fig. 4(g)], establishing these defects as half-monopoles, or boojums.

In the midplane, the polarization field exhibits a defect core with vanishing magnitude, around which the director field is dominated by bend rather than splay deformations [Fig. 4(h)]. The dominance of bend deformations arises from the minimization of electrostatic energy [52,59–63]. Splay deformations in the polarization field generate bound charge density ($\rho = -\nabla \cdot \mathbf{P}$), and the resulting Coulomb interactions between these charges increase the electrostatic energy, thereby penalizing splay. Consequently, bend distortions become energetically favorable in the vicinity of the defect core. The validity of the calculated polarization field is supported by its ability to quantitatively reproduce the spiral optical textures observed experimentally (Fig. S6 [41]).

The polarization field in planes located between the point defects exhibits a characteristic structure: at the core, the polarization is oriented perpendicular to the $xy$-plane, whereas toward the periphery it lies entirely within the plane [Fig. 4(i, j)]. This configuration corresponds to a $\pi/2$ rotation from the center to the periphery, a hallmark of meronic textures. Mapping the in-plane polarization onto the ground-state manifold shows that it covers either the upper or lower hemisphere of $\mathbb{S}^2$, thereby confirming the identification of these structures as merons.

Analysis of these three configurations reveals that topological charge is conserved during the defect evolution across the N–$\text{N}_\text{F}$ phase transition. To quantify this, consider a cylindrical surface enclosing the defects and oriented perpendicular to the cell substrates. On any intersecting plane parallel to the substrates, this cylinder outlines a closed loop. The winding number of the polarization field evaluated along this loop is invariant across all parallel planes and equals the net topological charge Q of the imprinted pattern, provided no disclination line passes through the lateral surface of the cylinder. Specifically, the Q = 0 configuration (the ±1/2 defect pair) enters the $\text{N}_\text{F}$ phase free of bulk singularities. Conversely, both configurations carrying Q = +1 (the +1/2 disclination pair and the +1 disclination) each host a monopole of unit topological charge.

*Conclusions*—In summary, we have demonstrated that inversion symmetry-breaking universally converts apolar disclinations into hybrid polar defects through two mechanisms: the topological prohibition of isolated half-integer disclinations and the topological instability of integer disclinations. Because disclinations emerge spontaneously during continuous symmetry-breaking at the I–N transition, these results establish a generalized Kibble mechanism for ferroelectric nematics, in which cascaded continuous and discrete symmetry breaking generates hybrid topological defects. The three hybrid configurations identified here are fundamental, though not exhaustive, topological motifs of the $\text{N}_\text{F}$ phase, and their intrinsic topological stability renders them ubiquitous across a range of conditions, as confirmed both by our experiments in untreated cells and by independent reports in the literature [28,34,64,65] (see End Matter).

Prescriptive control of hybrid defect topology via photopatterning provides a robust route for engineering polar textures on demand, opening opportunities in nonlinear light-matter interactions. Ferroelectric nematic fluids exhibit exceptionally large second-order nonlinear optical coefficients [66], offering new opportunities in nonlinear and quantum optics through their intrinsic tunability [67]. With dimensions comparable to optical wavelengths, hybrid defects can modulate the momentum of light, creating structured beams such as optical vortices, while simultaneously achieving second-harmonic generation [32,64,68].

*Acknowledgments*—We acknowledge the financial support by National Key Research and Development Program of China under grant 2022YFA1405000, National Natural Science Foundation of China under grants 12174177 and 12204226, Guangzhou Basic and Applied Basic Research Foundation under grant 2024B1515040023, and GJYC Program of Guangzhou via grant 2024D03J0002. R. Z. acknowledges support by Research Grants Council of Hong Kong via grant no. 16308225. We also thank Laurent Bellaiche, Lang Chen, Oleg D. Lavrentovich, Jonathan V. Selinger, Robin B. Selinger, and Slobodan Žumer for many valuable discussions.

*Contact author: weiqh@sustech.edu.cn.